# The Role of Mobile Point Defects in Two-Dimensional Memristive Devices

*Benjamin Spetzler\*, Dilara Abdel, Frank Schwierz, Martin Ziegler and Patricio Farrell*

B. Spetzler, F. Schwierz

Micro- and Nanoelectronic Systems, Department of Electrical Engineering and Information Technology, Technische Universität Ilmenau, Ehrenbergstraße 29, 98693 Ilmenau, Germany

M. Ziegler

Micro- and Nanoelectronic Systems, Department of Electrical Engineering and Information Technology, Technische Universität Ilmenau, Ehrenbergstraße 29, 98693 Ilmenau, Germany

Institute of Micro and Nanotechnologies MacroNano, Technische Universität Ilmenau, Ehrenbergstraße 29, 98693 Ilmenau, Germany

D. Abdel, P. Farrell

Numerical Methods for Innovative Semiconductor Devices, Weierstrass Institute for Applied Analysis and Stochastics (WIAS), Mohrenstraße 39, 10117 Berlin, Germany

\*Author to whom correspondence should be addressed. E-mail: benjamin.spetzler@tu-ilmenau.de

## Abstract

Two-dimensional (2D) layered transition metal dichalcogenides (TMDCs) are promising memristive materials for neuromorphic computing systems as they could solve the problem of the excessively high energy consumption of conventional von Neumann computer architectures. Despite extensive experimental work, the underlying switching mechanisms are still not understood, impeding progress in material and device functionality. This study reveals the dominant role of mobile defects in the switching dynamics of 2D TMDC materials. The switching process is governed by the formation and annihilation dynamics of a local vacancy depletion zone. Moreover, minor changes in the interface potential barriers cause fundamentally different device behavior previously thought to originate from multiple mechanisms. The key mechanisms are identified with a charge transport model for electrons, holes, and ionic point defects, including image-charge-induced Schottky barrier lowering (SBL). The model is validated by comparing simulations to measurements for various 2D $MoS_2$-based devices, strongly corroborating the relevance of vacancies in TMDC devices and offering a new perspective on the switching mechanisms. The insights gained from this study can be used to extend the functional behavior of 2D TMDC memristive devices in future neuromorphic computing applications.

**Keywords**: 2D materials, memristive devices, transition metal dichalcogenides, numerical modeling, drift-diffusion equations, Schottky barrier lowering

# 1 Introduction

Artificial intelligence (AI) is regarded as one of the most important emerging technologies of the 2020s and a key enabler for a variety of disruptive innovations. It will continue to have major implications on the worldwide economy. Even if not all optimistic expectations of the early days will be fulfilled, AI will likely change the way people live and work to a greater degree than the introduction of the personal computer, the internet, or the smartphone did in the past [1]. To unfold its full potential, AI requires suitable hardware, which, unfortunately, presents a major obstacle [2]. The conventional hardware for information processing, i.e., computers based on digital CMOS (complementary metal-oxide-semiconductor) logic and the von Neumann architecture, is far from ideal for AI applications. In particular, the energy consumption of information processing in AI systems based on this hardware is unacceptably high. We note that the energy consumption of AI is much more than a matter of merely academic interest. Instead, serious concerns exist about the rapidly growing carbon footprint of AI hardware and its harmful impact on the environment and the global climate. [3–5] For example, it is estimated that training ChatGPT may have consumed as much electric power as 175,000 Danes within a month [6].

An alternative, significantly more energy-efficient, and thus more sustainable solution for many AI problems is using hardware that processes information similarly to biological brains, i.e., neuromorphic computing [7,8]. Such hardware can be realized using memristive devices, for short memristors, as a fundamental building block. Memristors permit emulating the operation of synapses and neurons, the basic elements of biological brains, in an elegant way. Their current-voltage (I-V) characteristics are hysteretic; therefore, their resistance can be modulated by applying voltage pulses. So far, most memristors have been vertical two-terminal metal-insulator-metal structures [9,10]. These devices have been used in nonvolatile memory architectures, frequently referred to as RRAM (resistive random-access memory) or memristive RAM [11].

Recently, memristors made of 2D and quasi-2D layered materials have attracted considerable attention [12,13]. A promising class of layered materials for memristors are transition metal dichalcogenides (TMDCs) such as $MoS_2$, $MoSe_2$, $WS_2$, and $WSe_2$. Both vertical and lateral TMDC memristors have been demonstrated, each offering specific beneficial features. Vertical TMDC memristors comprise a layer sequence of a top contact (metal), active material (TMDC), and bottom contact (metal) [14–17]. They show excellent scaling potential, can easily be implemented in crossbar structures, and offer memristive switching at low voltages. For example, vertical electroforming-free $MoS_2$ memristors with switching voltage swings $\Delta U$ ($\Delta U = U_{set} - U_{reset}$, where $U_{set}$ is the set voltage and $U_{reset}$ is the reset voltage) in the range 160 - 550 mV have been reported [14–16], while the lowest reported voltage swings for traditional vertical MIM (metal-insulator-metal) memristor voltages are around/above 1 V [18,19]. Lateral TMDC memristors (Fig. 1), on the other hand, have a larger footprint and require higher switching voltages than their vertical counterparts but offer a higher degree of functionality since lateral structures can be equipped with additional terminals such as top or bottom gates. These gates permit adjusting the device properties and enable several desirable performance features such as improved device linearity and symmetry, increased number of resistance states, controllable synaptic weight, and complex neuromorphic learning [20–23]. Thus, lateral TMDC memristors, whether equipped with additional terminals or not, show promise for memristive neuromorphic systems for future AI solutions. Unfortunately, the physics of lateral TMDC memristors is not yet understood, and even the mechanisms responsible for resistive switching in these devices are still debated. Indications for three different switching mechanisms were found experimentally.

The first switching mechanism (Fig. 1a) is connected with the charging and discharging dynamics of trap states [24–30]. These states could be introduced by defects far away from the electrodes (in the bulk) or at the semiconductor-electrode interfaces [31,32]. They can capture and release electrons or holes with a probability depending on the applied voltage, thereby altering the bulk and interface electrostatic potential, free charge carrier density, and contact barriers [33]. Experimental investigations based on current-voltage measurements and analytical approximations resulted in estimated trap area densities between $\approx 10^{12}$ cm$^{-2}$ [24] and $\approx 10^{15}$ cm$^{-2}$ [27] for $MoS_2$. The second mechanism (Fig. 1b) is based on charged point defects sufficiently mobile to drift in the applied electric field. Experimental indications were found for the accumulation and drift of sulfur vacancies along grain boundaries in the conducting channel of lateral monolayer $MoS_2$ devices [34]. This observation is consistent with other experimental work [31,35] and was correlated with hysteresis in the current-voltage characteristics [34]. Elsewhere [36], measurements were performed on exfoliated $MoS_2$ multilayers, which were plasma treated to increase the concentration of sulfur vacancies. The homogeneous drift of mobile sulfur vacancies in the conduction channel was confirmed by local Auger-electron spectroscopy and experimentally correlated with hysteresis in the current-voltage characteristics [36]. Sulfur vacancies could behave as n-type dopants in $MoS_2$ [37,38], thereby changing the conductivity in the bulk and modifying the potential energy barrier at the semiconductor-metal interfaces. The third mechanism (Fig. 1c) is based on an electric-field-induced transition from a low-conductive semiconducting phase to a highly conductive metallic phase of the TMDC material. Such an effect has been demonstrated in lateral MoTe$_2$ devices, where a large electric field caused a transition from the semiconducting 2H phase to a metastable high-conductive 1T' metallic phase [39], and in 2H-Te$_2$ with a transition from the 2H phase to an unstable distorted 2H$_d$ phase [40]. In MoS$_2$, this effect was achieved indirectly by a pre-intercalation of Li$^+$ ions, which are highly mobile in between the monolayer sheets and induced the 2H to 1T' transition upon drift and agglomeration in the applied electric field [41]. Despite this experimental progress, the switching mechanism in 2D TMDC devices lack a profound physical understanding.

In particular, the first two mechanisms seem promising for application [42] but are superficially similar because the change in conductance is based on a change in the charge density in both cases. Further, the two mechanisms could be coupled and occur in parallel since point defects can introduce additional electronic states [43,44]. These similarities make distinguishing the two mechanisms challenging or even impossible with experimental correlation studies. While trap-state-based hysteresis is assumed in most work on lateral devices to explain resistive switching [12–18], it is unclear whether or not the drift of charged point defects could contribute as well or even dominate the hysteresis in some cases. Note that quantitative physical models could be powerful in separating such effects, which are experimentally difficult to access or distinguish. However, the few reported models of lateral devices are based on analytical approximations of the I-V curve with many fitting parameters [20] or compact-model formulations [45]. While such models can be helpful in circuit simulations and design, they do not capture the complex dynamics of the switching process.

In this work, we analyze the role of mobile point defects in the switching mechanism of 2D memristive devices and demonstrate the dominant impact they can have on the hysteresis and the device characteristics. A charge-transport model is introduced, solved via a finite-volume discretization, and compared to experimental data of lateral memristive devices based on exfoliated MoS$_2$. The model self-consistently solves semi-classical equations for the electrostatic potential and quasi-Fermi potentials of electrons, holes, and ionic sulfur vacancies. Additionally, we develop a mathematical formulation of image-charge-induced Schottky barrier lowering (SBL). This effect describes the change of Schottky barriers caused by an applied voltage and a redistribution of charge carriers in the semiconductor [46,47]. While SBL is often considered negligible [47], Kelvin probe force microscopy measurements on lateral devices demonstrated significant barrier-lowering effects [48]. Hence, SBL could lead to hysteresis by altering the interface-limited conduction. Despite its fundamental character, up to now, there is no mathematical formulation of this effect for charge-transport models, and the role of SBL in the hysteresis still remains unclear.

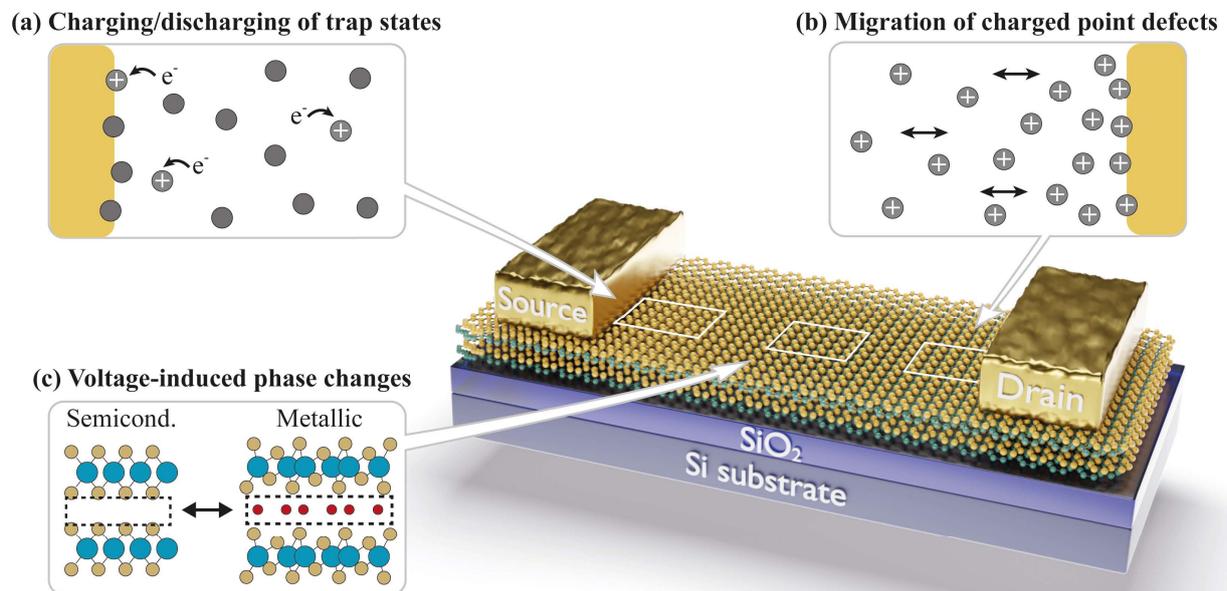

**Fig. 1** Example of a lateral memristive device structure comprising two electrodes (source and drain) on top of a two-dimensional memristive material. The three illustrated mechanisms have been suggested as the origin of hysteresis in lateral memristive devices based on 2D TMDC materials: **(a)** the charging and discharging dynamics of trap states [12–18], **(b)** the migration of charged point defects in the electric field [34,36], and **(c)** the voltage-induced phase change from a stable, low-conductive semiconducting phase to a metastable, high-conductive metallic phase [39,41].

## 2 Charge Transport Model

This section presents a semi-classical charge transport model. The equations and boundary conditions are detailed in the following, and the underlying assumptions are discussed. First, we present a system of equations without SBL in Subsections 2.1 and 2.2 before including SBL in Subsection 2.3.

### 2.1 Bulk Equations

Without the effect of SBL, the charge-transport model comprises a system of four coupled nonlinear partial differential equations: three drift-diffusion equations for the quasi-Fermi potentials $\varphi_\alpha(\mathbf{x}, t)$, $\alpha \in \{n, p, x\}$ of electrons n, holes p, and mobile defects x, as well as Poisson's equation for the electrostatic potential $\psi(\mathbf{x}, t)$. The corresponding densities of the charge carriers are denoted by $n_\alpha(\mathbf{x}, t), \alpha \in \{n, p, x\}$. For the location vector $\mathbf{x}$ and the time $t$ it is $(\mathbf{x}, t) \in \Omega \times [0, t_e]$ for the spatial domain of the semiconductor $\Omega \subset \mathbb{R}^d$ ($d = 1, 2, 3$) and some given end time $t_e > 0$.

**Continuity equations.** The electron, hole, and ionic defect densities $n_n$, $n_p$, and $n_x$ satisfy the conservation of mass via the continuity equations

$$q_\alpha \partial_t n_\alpha + \nabla \cdot \mathbf{j}_\alpha = q_\alpha r_\alpha, \quad \text{with} \quad \alpha \in \{n, p, x\}, \tag{1}$$

where the charge $q_\alpha$ of the respective carrier species $\alpha$ is given by

$$q_\alpha = z_\alpha q, \quad \text{with} \quad \alpha \in \{n, p, x\}, \tag{2}$$

with the positive elementary charge $q$. The charge numbers are $z_n = -1$ and $z_p = +1$ for electrons and holes, respectively, and $z_x \in \mathbb{Z}$ for mobile defects. The source term on the right-hand side of Eq. (1) includes the production/reduction rates $r_\alpha$, which can be used to consider recombination and generation processes of charge carriers. The current densities $\mathbf{j}_\alpha$ are proportional to the negative gradient $-\nabla \varphi_\alpha$ of their respective quasi-Fermi potentials $\varphi_\alpha$, and are given by

$$\mathbf{j}_\alpha = -z_\alpha^2 q \mu_\alpha n_\alpha \nabla \varphi_\alpha, \quad \text{with} \quad \alpha \in \{n, p, x\}, \tag{3}$$

with the charge carrier mobilities $\mu_\alpha$. The charge carrier densities $n_\alpha$ can be linked to the respective quasi-Fermi potentials $\varphi_\alpha$ and the electrostatic potential $\psi$ via the state equations [49]

$$n_\alpha(\psi, \varphi_\alpha) = N_\alpha \mathcal{F}_\alpha(\eta_\alpha), \quad \text{with} \quad \eta_\alpha = \frac{q_\alpha(\varphi_\alpha - \psi) + z_\alpha E_{\alpha,0}}{k_B T}, \text{ and } \alpha \in \{n, p, x\}, \tag{4}$$

with the temperature $T$, the Boltzmann constant $k_B$, and the effective densities of state of the conduction band and valence band $N_c = N_n$ and $N_v = N_p$, respectively. For the statistics functions $\mathcal{F}_n$ and $\mathcal{F}_p$ of electrons and holes, we use the Fermi-Dirac integral of order $1/2$ and for mobile defects the Fermi-Dirac integral of order $-1$, i.e., $\mathcal{F}_x(\eta) = 1/(\exp(-\eta) + 1)$. The latter function is selected because it provides an upper limit for $n_x$ and can be shown to be thermodynamically consistent [50]. Therefore, $N_x$ can be physically interpreted as the maximum possible density of mobile defects prescribed by the material, and $E_{x,0}$ as an intrinsic energy level of the defects [50]. Further, $E_{c,0} = E_{n,0}$ and $E_{v,0} = E_{p,0}$ denote the intrinsic band edge energies of the conduction and valence bands for electrons and holes, respectively. These energies relate to the electron affinity $\chi_e$ and the bandgap $E_g$ via $E_{c,0} := -\chi_e$ and $E_{v,0} := -\chi_e - E_g$. Consistent with Eq. (4), the electric potential dependent conduction band edge $E_c$, valence band edge $E_v$, and vacancy energy level $E_x$ are defined as

$$E_\alpha := q_n \psi + E_{\alpha,0} \quad \text{with} \quad \alpha \in \{c, v\}, \tag{5}$$

$$E_x := q_x \psi - z_x E_{x,0}. \tag{6}$$

**Poisson equation.** Finally, the three continuity equations (Eq. (1)) are self-consistently coupled via the electrostatic potential $\psi$ to the nonlinear Poisson equation

$$-\nabla \cdot (\varepsilon \nabla \psi) = q_C C + \sum_{\alpha \in \{n,p,x\}} q_\alpha n_\alpha(\psi, \varphi_\alpha). \tag{7}$$

In this equation, $C$ is either an effective acceptor or donor background dopant density with dopant charge $q_C = q z_C$, described by the background dopant charge number $z_C \in \{-1, 1\}$. The quasi-static electric permittivity $\varepsilon = \varepsilon_0 \varepsilon_r$ is expressed as the product of the electric vacuum permittivity $\varepsilon_0$ and the quasi-static relative electric permittivity $\varepsilon_r$.

## 2.2 Boundary Conditions at Metal-Semiconductor Contacts

The metal-semiconductor contacts $\Gamma$, are assumed to represent physical barriers for mobile defects, and therefore, we impose the zero-flux condition (for $x \in \Gamma$)

$$\boldsymbol{j}_x(\boldsymbol{x},t) \cdot \boldsymbol{v}(\boldsymbol{x}) = 0, \tag{8}$$

with the outward directed unit normal vector $\boldsymbol{v}$ of the semiconductor domain. For electrons and holes, thermionic emission is described with the flux boundary conditions [51]

$$\begin{aligned}\boldsymbol{j}_n(\boldsymbol{x},t) \cdot \boldsymbol{v}(\boldsymbol{x}) &= q_n v_n\bigl(n_n(\boldsymbol{x},t) - n_{n,0}\bigr), \\ \boldsymbol{j}_p(\boldsymbol{x},t) \cdot \boldsymbol{v}(\boldsymbol{x}) &= q_p v_p\bigl(n_p(\boldsymbol{x},t) - n_{p,0}\bigr).\end{aligned} \tag{9}$$

Here, $v_n$ and $v_p$ are the electron and hole recombination velocities given by (with $h$ as Planck constant)

$$v_n = \frac{4\pi m_n^*(k_B T)^2}{h^3 N_c}, \quad v_p = \frac{4\pi m_p^*(k_B T)^2}{h^3 N_v}, \tag{10}$$

with the effective mass $m_n^*$ and $m_p^*$ of electrons and holes, respectively. The corresponding quasi-equilibrium carrier densities $n_{n,0}$ and $n_{p,0}$ at the contacts are

$$n_{n,0} = N_c \mathcal{F}_n\left(-\frac{\phi_0}{k_B T}\right), \quad n_{p,0} = N_v \mathcal{F}_p\left(-\frac{E_g - \phi_0}{k_B T}\right), \tag{11}$$

where $\phi_0 = \phi_0(\boldsymbol{x}) > 0$ is an intrinsic Schottky energy barrier constant in time and $E_g = E_{c,0} - E_{v,0}$ the band gap. For the electrostatic potential at the metal-semiconductor contacts ($\boldsymbol{x} \in \Gamma$) we apply the Dirichlet condition

$$\psi(\boldsymbol{x},t) = \psi_0(\boldsymbol{x}) + \psi_a(\boldsymbol{x},t), \tag{12}$$

with $\psi_0 = (\phi_0 - E_{c,0})/q_n$ denoting an intrinsic electrostatic potential barrier, and $\psi_a$ denoting an applied electrostatic potential at the contact. The intrinsic Schottky barrier $\phi_0$ is considered as a material and device parameter, i.e., for a given device, $\psi_0$ and $\phi_0$ depend only on the location $\boldsymbol{x}$. Zero-flux conditions can be used at all non-contact boundaries.

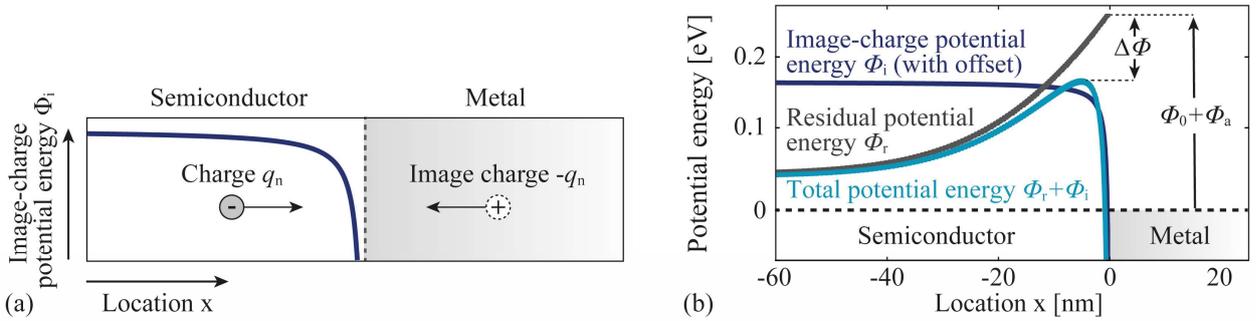

**Fig. 2** Schematic illustration of **(a)** an electron point charge $q_n$ at the semiconductor-metal interface inducing an image charge $-q_n$ in the metal electrode, which results in an attractive electrostatic potential energy $\phi_i$ (image-charge potential energy), and **(b)** illustration of the superposition of the residual potential energy $\phi_r$ and the image-charge potential energy $\phi_i$, which reduces the interfacial potential energy barrier maximum by $\Delta\phi$ relative to the value of $\phi_r$ at the interface. The boundary value of $\phi_r$ at the interface is prescribed by the sum of the intrinsic Schottky energy barrier $\phi_0$ and the applied potential energy $\phi_a$ ($\phi_a = q_n \psi_a$, with the applied electrostatic potential $\psi_a$). Note that an offset of approximately 0.16 eV is added to $\phi_i$ for illustrative purposes and $\phi_i = 0$ eV far away from the electrode, i.e., at $x \to -\infty$ in this example.

## 2.3 Image-Charge Induced Schottky Barrier Lowering

The two subsections above describe a system where the electrostatic potential at the boundary is prescribed by an intrinsic Schottky barrier and only the applied electrostatic potential $\psi_a$ adds a time-dependent contribution. However, any point charge present in the semiconductor induces a charge of opposite polarity (image charge) in the metal electrode, and this adds a contribution $\phi_i$ to the potential energy of the charge (image-charge potential energy $\phi_i$) with an energy minimum at the contact interface [51] as schematically illustrated in Fig. 2a. The image-charge potential energy $\phi_i$ superposes with the residual potential energy $\phi_r \coloneqq$ $q_\mathrm{n}\psi_\mathrm{r}$ ($\psi_\mathrm{r}$: residual potential), which corresponds to the potential energy without image charge. As a result, the barrier maximum of the total potential energy at the interface is reduced by a value $\Delta\phi$ compared to the case without image-charge contribution (Fig. 2b). While the image-charge potential energy is constant in time, $\phi_\mathrm{r}$ depends on the charge distribution within the semiconductor and the applied potential $\phi_\mathrm{a} \coloneqq q_\mathrm{n}\psi_\mathrm{a}$, which can correspondingly influence $\Delta\phi$.

Including image-charge-induced barrier lowering requires modifying the bulk equations as well as the boundary conditions. First, we modify the boundary condition for the electrostatic potential (Eq. (12)) at the metal-semiconductor contact ($\boldsymbol{x} \in \boldsymbol{\Gamma}$) by adding the change $\Delta\psi \coloneqq \Delta\phi/q_\mathrm{n}$ of the potential barrier induced by the image charge. The new boundary condition reads

$$\psi(\boldsymbol{x},t) = \psi_0 + \psi_\mathrm{a}(\boldsymbol{x},t) + \Delta\psi(\boldsymbol{x},t), \quad \boldsymbol{x} \in \boldsymbol{\Gamma}. \tag{13}$$

Within the classical approximation [51], the change $\Delta\psi$ of the potential barrier induced by the image charge is only defined for an upwards-bent conduction band edge ($\nabla_v\psi_\mathrm{r} < 0$, $\boldsymbol{x} \in \boldsymbol{\Gamma}$) and can be expressed as a function of the projected gradient of the residual electrostatic potential $\nabla_v\psi_\mathrm{r} \coloneqq \nabla\psi_\mathrm{r} \cdot \boldsymbol{v}$, ($\boldsymbol{x} \in \boldsymbol{\Gamma}$),

$$\Delta\psi = \begin{cases} \sqrt{-\dfrac{q\nabla_v\psi_\mathrm{r}}{4\pi\varepsilon_\mathrm{i}}}, & \nabla_v\psi_\mathrm{r} < 0 \text{ (upwards bent } E_\mathrm{c}\text{)}, \\ 0, & \nabla_v\psi_\mathrm{r} \geq 0 \text{ (constant and downwards bent } E_\mathrm{c}\text{).} \end{cases} \tag{14}$$

In this equation, we refer to $\varepsilon_\mathrm{i} \coloneqq \mathrm{Re}\{\varepsilon^*(f_\mathrm{i})\}$ as the image-charge dielectric constant, defined as the real part $\mathrm{Re}\{\varepsilon^*(f_\mathrm{i})\}$ of the complex dielectric permittivity $\varepsilon^*$ at the frequency $f_\mathrm{i} = 1/T_\mathrm{i}$, which is connected to the time $T_\mathrm{i}$ an electron needs to travel from the barrier maximum to the metal-semiconductor interface (see Methods). The residual electrostatic potential $\psi_\mathrm{r}$ is defined as the electrostatic potential obtained as a solution to Poisson's equation

$$-\nabla \cdot (\varepsilon\nabla\psi_\mathrm{r}) = q_\mathrm{C}C + \sum_{\alpha\in\{\mathrm{n,p,x}\}} q_\alpha n_\alpha(\psi,\varphi_\alpha), \quad \boldsymbol{x} \in \boldsymbol{\Omega}, \tag{15}$$

which is now linear in $\psi_\mathrm{r}$ because $n_\alpha$ is a function of $\psi$. At the contact boundaries, $\psi_\mathrm{r}$ is obtained using the Schottky contact boundary condition

$$\psi_\mathrm{r}(\boldsymbol{x},t) = \psi_0(\boldsymbol{x}) + \psi_\mathrm{a}(\boldsymbol{x},t), \quad \boldsymbol{x} \in \boldsymbol{\Gamma}. \tag{16}$$

Further, the equations for the equilibrium charge carrier densities at the contacts ($\boldsymbol{x} \in \boldsymbol{\Gamma}$) in Eq. (11) must be altered to

$$n_\mathrm{n,0} = N_\mathrm{c}\mathcal{F}_\mathrm{n}\left(-\frac{\phi_0 + \Delta\phi}{k_\mathrm{B}T}\right), \quad n_\mathrm{p,0} = N_\mathrm{v}\mathcal{F}_\mathrm{p}\left(-\frac{E_\mathrm{g} - (\phi_0 + \Delta\phi)}{k_\mathrm{B}T}\right), \tag{17}$$

to consider the change of the barriers in equilibrium. If the effect of SBL is considered, we consequently solve a system of five coupled partial differential equations (Eq. (1), Eq. (7), and Eq. (15)) for five solution fields, namely the three quasi-Fermi potentials $\varphi_\mathrm{n}, \varphi_\mathrm{p}, \varphi_\mathrm{x}$, and the two electrostatic potentials $\psi$ and $\psi_\mathrm{r}$. The equation system is supplemented with the equilibrium solution as initial conditions.

## 3 Application to Lateral Memristive Devices

We implemented all equations and numerical examples in the Julia programming language to extend the software package ChargeTransport.jl [50]. It was developed to simulate charge transport in semiconductors and is based on a finite volume method for spatial discretization [52,53]. In the following, the equations from Section 2 are applied to model representative examples of lateral TMDC-based memristive devices that allow us to gain general insights into the hysteresis mechanism.

### 3.1 Model Geometry and Boundary Conditions

An illustration of the geometry of a typical lateral memristive device is shown in Fig. 3a, along with its cross section in the x-y plane in Fig. 3b. The devices comprise a TMDC flake on a SiO$_2$/Si substrate, and the flake is laterally sandwiched by two Ti/Au electrodes. Typical TMDC flakes in lateral memristive devices have a thickness $D$ of some nanometers and a width $W$ and length $L$ of the order of a few micrometers (e.g., [29,32,39,54–56]). We further simplify the model geometry and represent it by a one-dimensional (1D) conducting channel (Fig. 1c) defined by the flake length $L$ between the electrodes and the electrode contact area $A = DW$. Hence, the 1D model comprises only the semiconductor (TMDC) domain of length $L$ with two metal-semiconductor contacts at the locations $x_1$ and $x_2$, where we apply the corresponding boundary conditions. For the applied potential, we assume $\psi_\mathrm{a}(x_1) = 0$ V at the left contact and $\psi_\mathrm{a}(x_2) = \psi_\mathrm{a}(x_1) +$ $U(t)$ at the right contact. The voltage $U(t)$ is a piecewise linear interpolation with four support points within the period of length $t_c$ and the extrema $U_{max}$ and $-U_{max}$ (Fig. 1d). The source term in Eq. (1) is set to $r_\alpha = 0$ for all simulations. This is justified by the large electron and hole mobilities (section below) and the small voltage sweep rates of the order of some seconds per volt [36].

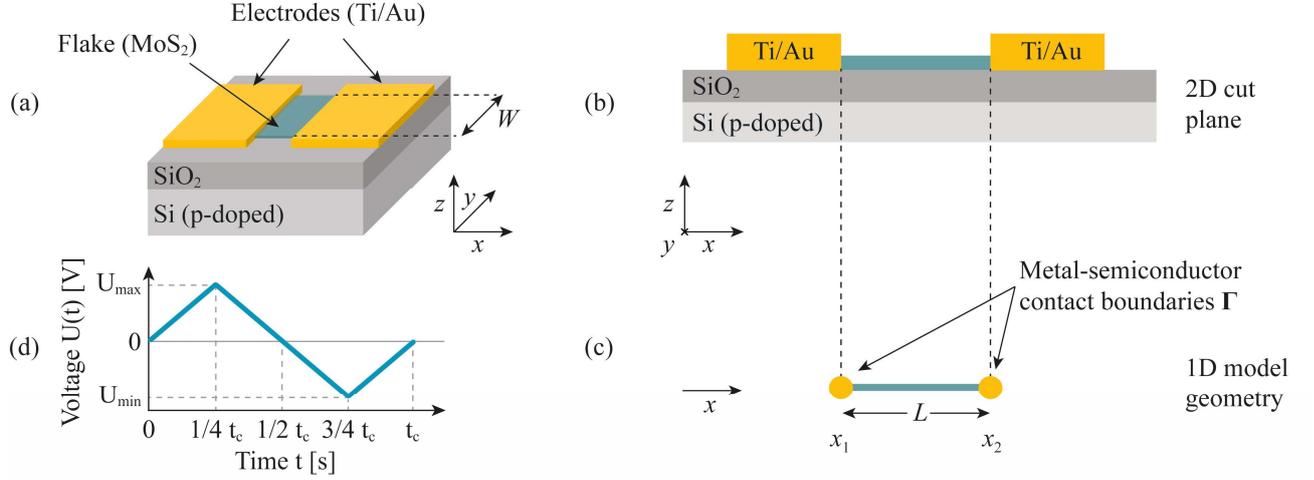

**Fig. 3** Illustration of **(a)** the geometry of the modeled device in three dimensions with indicated SiO$_2$/Si substrate, TMDC flake, and Ti/Au electrodes. **(b)** 2D view of the x-z cut plane of the memristive device, and **(c)** simplification of the geometry for the one-dimensional (1D) model with indicated channel length $L$, and locations $x_1$ and $x_2$ of the two contact boundaries $\Gamma$. **(d)** Illustration of the piecewise linear voltage $U(t)$ as a function of the time $t$ over a period of length $t_c$, and with a maximum voltage of $U_{max}$ and a minimum voltage $U_{min} = -U_{max}$.

## 3.2 Material Parameters

To gain general insights into the hysteresis mechanism and thoroughly validate the model, we use MoS$_2$ as a well-known example material for the TMDC domain. Auger-electron spectroscopy measurements revealed a large density of sulfur vacancies sufficiently mobile to drift under an applied electric field in this material [36]. The vacancies can be introduced during the sample fabrication with a plasma treatment step [36]. Because of the formation energy of sulfur vacancies of up to several eV depending on the conditions [43,44,57], we assume that the number of sulfur vacancies remains constant during the operation of the device. While the role of sulfur vacancies for doping and their stable charge states have been debated [43,58,59], for large concentrations, they are expected to behave as n-type dopants [58]. Therefore, we assume that the mobile ionic species x represents donor-type sulfur vacancies, single-positively charged with $z_x = +1$ in their ionized state. The maximum density of vacancies $N_x$ is estimated from the density of sulfur sites to $N_x \approx 4/V \approx 4 \cdot 10^{28}$ m$^{-3}$ (see Methods).

The electronic properties of MoS$_2$ are well investigated and depend significantly on the number of monolayers and other extrinsic and intrinsic factors, such as the stress state and the temperature [37,60,61]. In the following, we review and estimate the range of material parameters expected for the most stable phase 2H-MoS$_2$, a layer thickness of $D = 10 - 15$ nm at a temperature of $T = 300$ K.

With increasing thickness, the conduction band minimum (CBM) and the valence band maximum (VBM) shift from the K point (in the monolayer) to a midpoint between $\Gamma$ and K (CBM) and the $\Gamma$ point (VBM) (in the bulk) [62]. A transformation occurs from a direct semiconductor to an indirect one [54,63–65], with a gradual shift of the electronic bandgap from $E_g = 1.8$ eV (monolayer) to $E_g = 1.2$ eV (bulk). This shift has been explained theoretically by an increasing contribution of interfacial stress to the overall atomic configuration as the material gets thinner [66] and is consistent with *ab initio* calculations for the influence of stress on the electronic properties of MoS$_2$ monolayers [67,68]. Assuming no external or internal strain of the MoS$_2$ layer, e.g., from the fabrication process [69] or defect clusters [59], simulations predict values of $E_g \approx$ 1.3 eV for a layer thickness of 15 nm [66]. The changing band structure with layer thickness also alters the effective masses $m_n^*$ and $m_p^*$ of electrons and holes. For 2H-MoS$_2$ monolayers, values at the K point of $m_n^* = (0.45 - 0.57) m_0$ and $m_p^* = (0.59 - 0.64) m_0$ are reported [62,68,70], with geometric means of $\bar{m}_n^* \approx 0.54 m_0$ and $\bar{m}_p^* \approx 0.61 m_0$ [70] considering the slight anisotropy ($m_0$: electron rest mass). This is slightly smaller than the effective masses of the bulk with values of $m_n^* \approx 0.55 m_0$ and $m_p^* \approx 0.71 m_0$ [62]. For a layer thickness of 15 nm, we expect effective masses close to the values in the bulk, and use them to estimate the effective density of states $N_c$ and $N_v$ of electrons and holes with the three-dimensional parabolic band approximation ($\hbar$: reduced Planck constant)

$$N_c = 2\left(\frac{m_n^* k_B T}{2\pi\hbar^2}\right)^{3/2}, \quad N_c = 2\left(\frac{m_p^* k_B T}{2\pi\hbar^2}\right)^{3/2}. \tag{18}$$

Experimental and theoretical studies on the in-plane quasi-static dielectric permittivity of monolayer MoS$_2$ report a large scatter of values of $\varepsilon_r \approx 3.7$ [71–73], and $\varepsilon_r \approx 10 - 17$ [60]. Generally, $\varepsilon_r$ tends to increase as a function of the film thickness and values of $\varepsilon_r \approx 10$ were measured for layer thicknesses of $D = 10 - 15$ nm (close to $\varepsilon_r \approx 12$ in the bulk) [74]. While these results are in excellent agreement with *ab initio* simulations [75], other calculations yield smaller values of $\varepsilon_r = 7.1$ for the bulk [73]. For the image-charge dielectric constant, the frequency dependency of the complex electric permittivity becomes relevant, but our estimation shows $\varepsilon_i \approx \varepsilon_r$ as a reasonable approximation (see Methods). Further, electron affinities of $\chi_e \approx 3.7$ eV [76], $\chi_e \approx 4.0$ eV [77,78] and $\chi_e \approx 4.3$ eV [79,80] are reported for 2H-MoS$_2$. A summary of the parameters and their estimated range is provided in Table 2 (see Methods).

While the abovementioned parameters are obtained directly from the literature and used for all simulations, other parameters are expected to be highly sensitive to the sample's microstructure, defect composition, and interface quality. These sample-dependent model parameters include the mobilities $\mu_\alpha$, the intrinsic Schottky barriers $\phi_0$ and the defect energy level $E_{x,0}$ or the corresponding vacancy density $n_x$. For the electron and hole mobilities, a large range of $\mu_{n,p} \approx 0.1 - 200$ cm$^2$/(Vs) [37,69,81–83] is reported. This range has been explained by the scattering of charge carriers with interfacial Coulomb impurities [81]. As a result, the mobilities tend to increase with the layer thickness and depend significantly on the gate dielectric and interface quality of the respective device [69,81]. *Ab-initio* studies find activation energies of diffusion of $E_a \approx 0.8 - 2.5$ eV for sulfur vacancies in MoS$_2$, depending strongly on the spatial configuration of the point defects [57]. This energy barrier is expected to be reduced along grain boundaries or interfaces [34], and a previous estimation based on Auger-electron spectroscopy measurements resulted in a much smaller value of $E_a = 0.6$ eV [36]. These measurements were performed on lateral memristive devices with vacancy densities of approximately $n_x = 10^{26} - 10^{27}$ m$^{-3}$ [36], assuming $N_x = 4 \cdot 10^{28}$ m$^{-3}$ (see Methods). With a work function of approximately 4.3 eV for Ti [84], slightly rectifying Schottky barriers around 0.3 eV are expected at the MoS$_2$/Ti contact according to the Schottky-Mott rule [36]. However, experimental values are smaller, with $\approx 0.05$ eV for Ti-MoS$_2$ contacts in equilibrium [85].

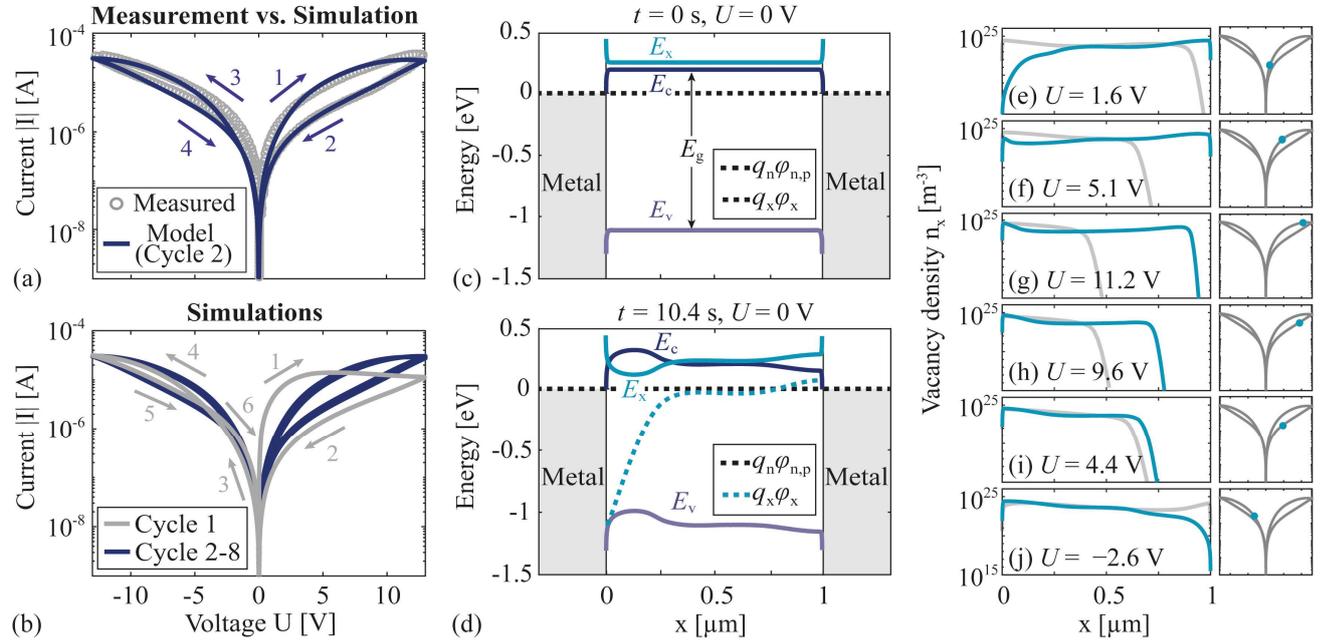

**Fig. 4** Simulations using the parameter set S$_1$ (Table 4, Methods). **(a)** Comparison of measurements with simulations of the second I-V cycle. **(b)** All eight consecutively simulated current-voltage cycles. **(c)** Simulated band diagram of the initial equilibrium configuration at $t = 0$ s, and an applied voltage $U = 0$ V, and **(d)** of the nonequilibrium configuration at the beginning of the second cycle at $t = 10.4$ s ($U = 0$ V). **(e-j)** Example configurations of the vacancy density $n_x(x)$ as a function of space $x$ for various positions in the I-V curve shown in (a), together with the quasi-static equilibrium configuration of $n_x$ (gray), i.e., the configuration that would be reached at a constant applied voltage for $t \to \infty$. The measured data are taken from [36].

# 4   Results

In the following, we analyze the origin of hysteresis, the influence of vacancy dynamics, and image-charge-induced Schottky barrier lowering in general for lateral memristive devices. We use parameters within the physically realistic range of MoS$_2$ (Section 3) as representative material and discuss the cases of vanishing Schottky barriers and significant Schottky barrier heights. Finally, we compare additional simulations with various experimental data sets to demonstrate the model's validity.

For the first case, we consider a parameter set S$_1$ (Table 1, Methods) with omittable Schottky barriers $\phi_0(x_1) = \phi_0(x_2) = 1$ meV. The experimental data used to obtain the sample-dependent model parameters are taken from [36] and are shown in Fig. 4a together with the simulation. The current magnitude $|I|$ is a smooth function of the applied voltage $U$ without visible discontinuities and with approximately equal maximum current magnitudes at $U = 13$ V and $U = -13$ V. While $|I(U)|$ is overall symmetric around $U = 0$ V, the right hysteresis branch ($U > 0$ V) has a slightly larger area compared to the left hysteresis branch ($U < 0$ V). As indicated, the direction of the hysteresis is clockwise and counterclockwise in the right and left hysteresis branch, respectively. The simulations of eight consecutive voltage cycles are shown in Fig. 4b, revealing a significant difference in the I-V characteristics of the first cycle compared to the following cycles. In particular, the I-V curve of the first cycle is highly asymmetric, with an initial rise in the current (Fig. 4b,1), which quickly saturates and almost remains constant until the maximum voltage is reached. In the left hysteresis branch (Fig. 4b, $U < 0$ V) a crossing of the current is visible, resulting in a change of the hysteresis direction from clockwise (Fig. 4b, 3-6) to counterclockwise Fig. 4b, 4-5). The other voltage cycles (Fig. 4b, Cycle 2-8) show a different but highly reproducible I-V behavior and reflect the characteristics of the measured I-V curve mentioned above very well, including the hysteresis direction and the transitions from the low-resistive state (LRS) to the high-resistive state (HRS) and vice versa.

The origin of the two I-V characteristics is the different initial configurations of the system at the beginning of the respective cycle. At the beginning of the first cycle ($t = 0$ s, $U = 0$ V), the modeled system is in equilibrium, i.e., all quasi-Fermi levels are constant $q_n \varphi_n = q_n \varphi_p = q_x \varphi_x = 0$ eV, as shown in the reduced band diagram (Fig. 4c). During the first cycle and at the beginning of the second cycle ($t = 10.4$ s, $U = 0$ V) the electrons and holes remain close to their quasi-static equilibrium configuration, i.e., the configuration that would be reached at a constant applied voltage for $t \to \infty$. The small mobility of vacancies on the other hand results in a substantial deviation from equilibrium (Fig. 4d). This deviation from equilibrium is apparent from the notable drop in the vacancy quasi-Fermi level $q_x \varphi_x$ at the left metal-semiconductor contact around $x = 0$. Such a nonequilibrium configuration is reproducibly reached after each cycle following the first one. This phenomenon appears to be superficially similar to electroforming in filamentary devices. There, the formation of a filament is induced during the initial voltage cycles and then modulated in the subsequent cycles [86]. However, in our model, vacancies drift homogeneously distributed over the cross section of the conducting channel, and we will show in the following that the difference between the first and subsequent hysteresis loops is just a special case of a more general dynamic phenomenon. In the following, we focus the discussion on the second cycle as a more reproducible and representative device characteristic.

In Fig. 4e-j, the vacancy density is shown at various voltages in the second hysteresis cycle. Additionally, its quasi-static equilibrium configuration at the respective voltage is shown (gray, faint). At the beginning of the cycle, the drop in $q_x \varphi_x$ (Fig. 4d) results in a local depletion of vacancies at the left electrode ($n_x \ll 10^{10}$ m$^{-3}$, Fig. 4e). This depletion zone initially limits the current, but it vanishes quickly as the voltage increases (Fig. 4e-f). Therefore, the device is in a high-conductance (LRS) state at the beginning of the second cycle. As the voltage is further ramped up, the current increases continuously until a depletion zone forms and grows at the right electrode (Fig. 4g). It limits the current just before the maximum voltage is reached. The depletion zone continues growing at the right electrode and reduces the current as the voltage is ramped down (Fig. 4g-i). Hence, the device is in the HRS. At $U = 0$ V, the system strives to restore equilibrium where no depletion zone is present. Because of the small vacancy mobility, the depletion zone is not entirely annihilated at the beginning of the left hysteresis branch (Fig. 4j, $U < 0$ V) leading to a vacancy density configuration similar to the one at the beginning of the right cycle. Therefore, the I-V curve is overall very symmetric around $U = 0$ V and the device starts in a high conductance state (LRS) at the beginning of the left hysteresis branch.

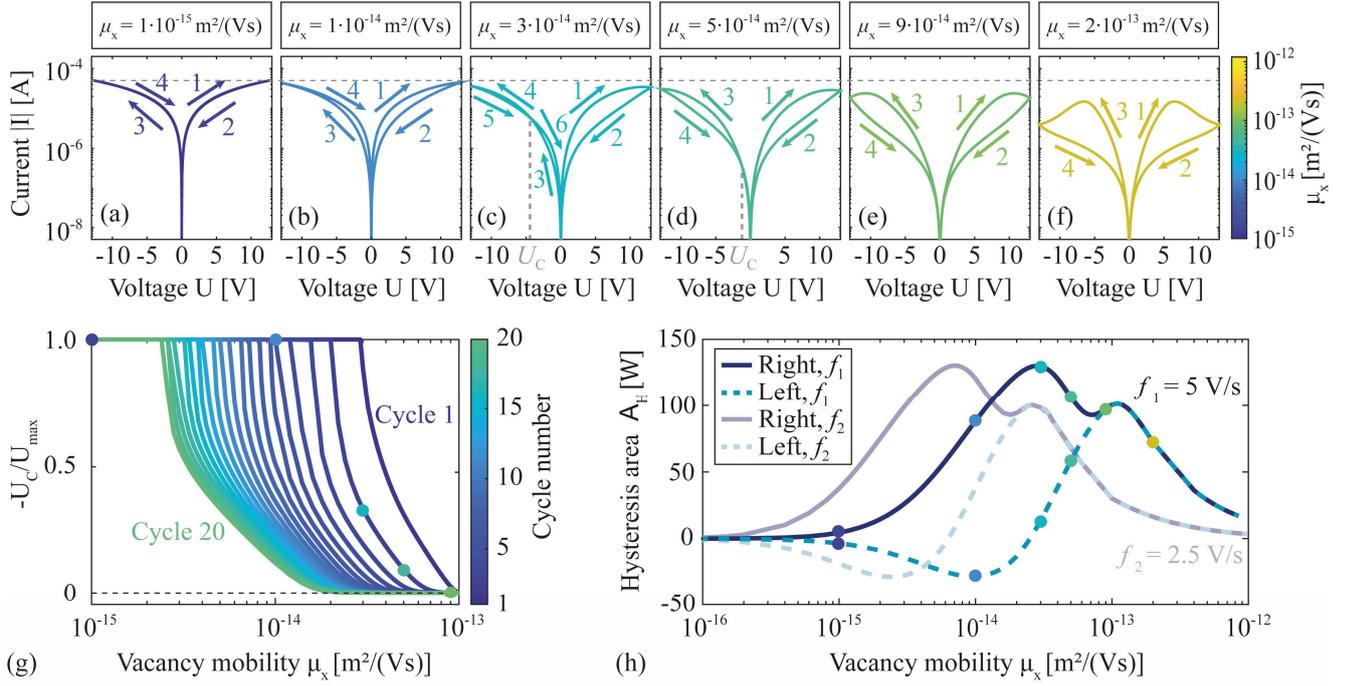

**Fig. 5** Simulation results illustrating the dependency of the I-V characteristics on the vacancy mobility $\mu_x$ and the voltage sweep rate $f$ using the parameter set $S_1$ (Table 4, Methods) **(a-f)** Example hysteresis loops of the second cycle for various vacancy mobility values simulated with the voltage sweep rate $f_1 = 5$ V/s. **(g)** Negative cross-over voltage $-U_c$ normalized to the maximum voltage $U_{max} = 13$ V for the first twenty voltage cycles as functions of $\mu_x$. The values from the second-cycle hysteresis curves in (a-e) are indicated with dots. **(d)** Hysteresis areas $A_H$ defined as the integrals over the right (bold) and left (dashed) hysteresis branch of the second sweep cycle for two different voltage sweep rates $f_1 = 5$ V/s and $f_2 = 2.5$ V/s. The values from the hysteresis curves in (a-f) are indicated with dots.

To analyze the dependency of the I-V curve on the vacancy mobility in detail, we performed simulations using the same parameter set $S_1$ (Table 4, Methods) as before but with different values for the vacancy mobility, ranging from $\mu_x = 1 \cdot 10^{-16}$ m²/(Vs) to $\mu_x = 1 \cdot 10^{-1}$ m²/(Vs) for two different voltage sweep rates $f_1 = 5$ V/s and $f_2 = 2.5$ V/s. We define the hysteresis areas $A_H$ of the left and right hysteresis branches as the integral over the respective branch and consider the sign of the integral to distinguish the direction of the hysteresis. The hysteresis areas of the second cycle are plotted in Fig. 5h as functions of $\mu_x$ for $f_1$ and $f_2$. Example values of $A_H$ calculated from the hysteresis curves in Fig. 5a-f are indicated. As expected, $A_H = 0$ for $\mu_x \to 0$ and $\mu_x \to \infty$ because the vacancy density either remains stationary or in quasi-static equilibrium. A significant hysteresis occurs only in the regime between approximately $\mu_x = 10^{-15}$ m²/(Vs) and $\mu_x = 10^{-12}$ m²/(Vs).

As the mobility exceeds small values $\mu_x \leq 1 \cdot 10^{-15}$ m²/(Vs), the hysteresis starts to open ($A_H \approx 0$) and is initially clockwise oriented in both branches (Fig. 5a-b). Because the depletion zone is still present when $U < 0$ V is reached the device is in the low conducting state at the beginning of the left hysteresis branch. A transition of the hysteresis direction occurs in the left hysteresis branch from clockwise ($A_H < 0$) to counterclockwise ($A_H > 0$) at intermediate mobilities around $\mu_x \approx 3 \cdot 10^{-14}$ m²/(Vs) (Fig. 5c). During the transition, the current in the left hysteresis branch intersects at a voltage $U_c$ (Fig. 5c, indicated), which continuously moves towards $U = 0$ V as the mobility increases until the entire hysteresis branch is directed counterclockwise (Fig. 5d). Origin of this transition is the shorter time required for vacancies with a larger mobility to reach quasi-static equilibrium, which causes a faster annihilation of the depletion zone before $U < 0$ V is reached. Further increasing the vacancy mobility to large values (Fig. 5d-f) results in the formation of distinct maxima at voltages $U < U_{max}$ (and $U > U_{min}$), an increasingly symmetric I-V curve, and an overall reduced maximum current. This behavior reflects the formation dynamics of the depletion zone. In the high-mobility regime, the depletion zone can form before the maximum positive or negative voltage is reached, which results in a drop in the current, leading to the formation of the maxima. Simultaneously, the vacancy density is close to its quasi-static equilibrium configuration, which results in the pronounced symmetry around $U = 0$ V and $U_c \approx 0$ V. The dependency of $-U_c/U_{max}$ as a function of the vacancy mobility is detailed in Fig. 5g for the first 20 voltage cycles as a measure for the symmetry of the I-V curve. Over the entire range of $\mu_x$, the voltage $U_c$ always reaches $U_c = 0$ after a sufficiently large number of voltage cycles. While > 20 cycles are required to reach the symmetric I-V curve for $\mu_x < 2 \cdot 10^{-14}$ m$^2$/(Vs), one to two cycles are enough for values close to $\mu_x = 1 \cdot 10^{-13}$ m$^2$/(Vs).

The results show that the hysteresis and its symmetry are governed by the formation and annihilation dynamics of the vacancy depletion zone, i.e., the deviation of the vacancy density from its quasi-static equilibrium configuration. Hysteresis occurs in a small vacancy mobility range where the vacancies are sufficiently mobile to follow the potential gradient but slow enough to be in nonequilibrium. Therefore, a change in the voltage sweep rate only shifts the $A_H(\mu_x)$ curve on the vacancy mobility axis (Fig. 5h, $f_2$). As another consequence, the asymmetry and intersection in the left hysteresis branch are volatile features. They depend on vacancy mobility and the cycle number and eventually vanish after enough sweep cycles. In this first I-V curve example, the intrinsic Schottky barriers are so small that SBL is negligible over the entire voltage range. The influence of Schottky barriers and SBL on the I-V curve is explored next in the second example.

For the second case, we consider a similar parameter set $S_2$ (Table 4, Methods) as before but introduce intrinsic Schottky barriers $\phi_0(x_1) = 0.144$ eV and $\phi_0(x_2) = 0.11$ eV with a slightly smaller value at the right contact. Fig. 4a demonstrates the excellent match of the simulated and measured I-V curve. Both have the same clockwise direction (right branch) and counterclockwise direction (left branch) as in Fig. 4a, but the asymmetry around $U = 0$ V is more pronounced. In contrast to the previous I-V curve (Fig. 4a), the two maximum current magnitudes are notably different, and the right hysteresis area is smaller than the left one. Because the intrinsic Schottky barriers are small but significant, SBL can occur. In equilibrium at $t = 0$ (Fig. 6b), the SBL notably reduces the contact barriers by approximately 25 % and 18 % to values of $\phi(x_1) \approx 0.11$ eV and $\phi(x_2) \approx 0.09$ eV. This difference in the reduction of the Schottky barriers is caused by the difference in the intrinsic Schottky barrier $\phi_0(x_1)$ and $\phi_0(x_2)$ at the contacts. At a large intrinsic Schottky barrier, the magnitude of the residual potential gradient $|\nabla_v \psi_r|$ tends to be larger than at a small barrier. This gradient determines the reduction of the Schottky barrier via Eq. (14), and therefore, SBL tends to reduce the difference in the Schottky barriers compared to the intrinsic barriers.

This affects the band edges $E_c$, $E_v$, and $E_x$ in proximity of the contact interfaces, correspondingly (Fig. 6b, $E_v$ and $E_x$ not shown). In Fig. 6d, the applied voltage $U$, the current magnitude $|I|$, and the change $\Delta\phi(x_1)$ and $\Delta\phi(x_2)$ of the Schottky barriers are plotted as functions of the time $t$ over the first two voltage cycles. As the voltage increases ($U > 0$ V) the band edges at the left contact remain bent upwards ($\nabla_v \psi_r < 0$), which results in an increased Schottky barrier $\Delta\phi(x_1) > 0$ (Eq. (14)). At the right contact, the band edges move to lower energies, and the projected electrostatic potential quickly reaches positive values ($\nabla_v \psi_r \geq 0$) indicating a downwards bent conduction band. As a result, the change in the right Schottky barrier remains zero, i.e., $\Delta\phi(x_2) = 0$, over most of the right hysteresis cycle ($U > 0\ V$) and does not limit the current. As the voltage approaches negative values, the situation reverses and the change in Schottky barriers is $\Delta\phi(x_1) = 0$ and $\Delta\phi(x_2) > 0$ over most of the left hysteresis cycle ($U < 0\ V$). Note that both $\Delta\phi(t)$ are asymmetric around the time at which the respective maximum positive and negative voltage is applied. This asymmetry is also present in the current magnitude and reflects the hysteresis in the time domain.

To identify the influence of SBL on the I-V characteristics, we performed a simulation with the same parameters as for the curve in Fig. 6a but without SBL. The two I-V curves with and without SBL are compared in Fig. 6c, and qualitative differences are apparent. Without SBL, the hysteresis direction of the right branch is reversed, and a pronounced asymmetry is visible around $U = 0$ V, i.e., largely different maximum currents and hysteresis areas of the left and right branches. This asymmetry is caused by a larger difference in the equilibrium Schottky barriers if SBL is not considered. The total hysteresis area without SBL is similar to the case with SBL. An additional I-V simulation was performed without SBL using the equilibrium Schottky barriers $\phi(x)$ obtained from the simulation with SBL as intrinsic Schottky barriers $\phi_0$ (Fig. 6c, No SBL, reduced $\phi_0$). The resulting I-V curve matches the one with SBL. This comparison shows that the hysteresis is not dominated by a change in the interfacial current but a limitation in the bulk conductance via the formation of the depletion zone. Consequently, SBL greatly affects the I-V curve, but it does not contribute significantly to the hysteresis area in this example.

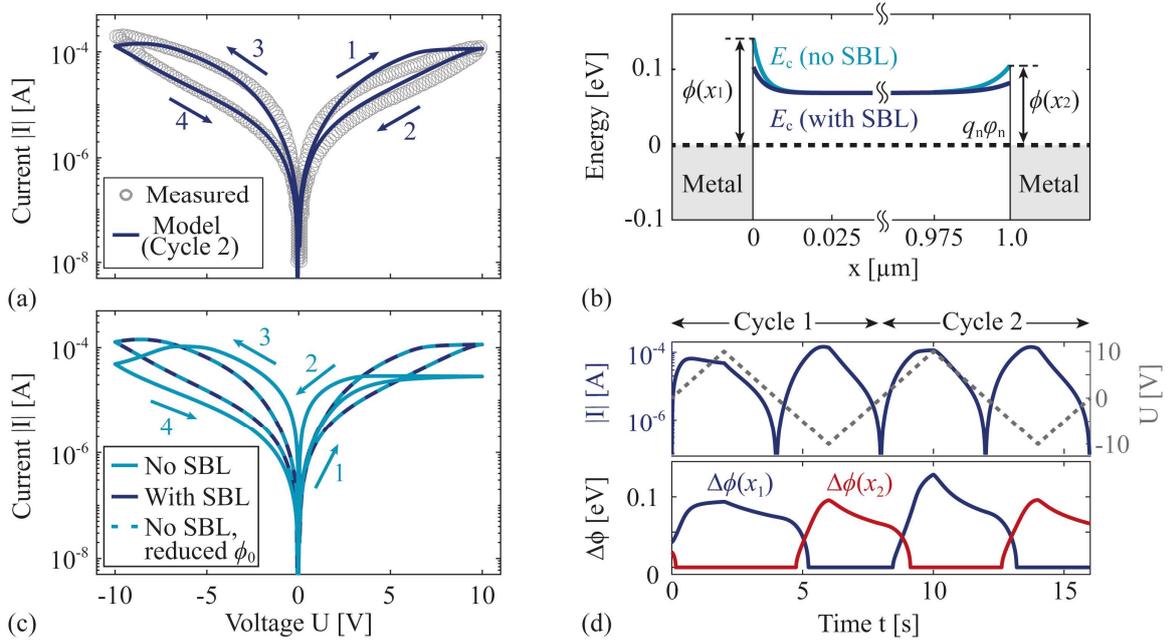

**Fig. 6** Comparison of simulations with measurements at the example of an asymmetric hysteresis curve using the parameter set $S_2$ (Table 4, Methods). **(a)** Simulation of the second voltage cycle and measurements. **(b)** Conduction band edges, and quasi-Fermi levels in equilibrium at $t = 0$ s with and without SBL. **(c)** Comparison of simulations of the second voltage cycle with and without SBL and without SBL but with intrinsic Schottky barriers equal to the equilibrium Schottky barriers from the case with SBL (No SBL, reduced $\phi_0$). **(d)** Magnitude of the current, applied voltage $U(t)$, and change of the Schottky barrier $\Delta\phi(t)$ as functions of the time $t$. The measured data are taken from [36].

The effect of a change in the Schottky barriers is illustrated with additional simulations in Fig. **7**, including SBL. We used the parameter set $S_2$ (Table 4, Methods) as before, but various intrinsic Schottky barriers for the right contact (Fig. 7a) and the left contact (Fig. 7b) with values from 0.12 eV to 0.18 eV while keeping the respective other Schottky barrier fixed to a value of 0.11 eV. In both cases, we observe a transition from a relatively symmetric I-V curve for the case of similar Schottky barriers to a notably asymmetric curve for significantly different Schottky barriers. This transition is accompanied by a change of the hysteresis direction in the left hysteresis branch if $\phi_0(x_1) < \phi_0(x_2)$ and in the right hysteresis branch if $\phi_0(x_2) < \phi_0(x_1)$. A comparison of the I-V curve in Fig. **7**b (right, $\phi_0(x_2) = 0.18$ eV) with the hysteresis curve in Fig. **6**c demonstrates that the behavior observed without SBL is mainly caused by the difference in the two Schottky barriers of the left and right contact. Other features, such as the formation of a maximum at $U < U_{\max}$ and the reduction of the maximum current magnitude with increasing barrier difference are also observed (Fig. **7**). These features are superficially similar to those caused by the increasing vacancy mobility of the device with symmetric contacts, but their origin differs. The asymmetric Schottky barriers result in different restoring forces compared to the symmetric configuration, which alters the formation and annihilation dynamics of the depletion zone.

To analyze the I-V dynamics caused by asymmetric Schottky contacts, we simulate the I-V curve of an example configuration over 100 consecutive voltage cycles (Fig. 7c). An asymmetry is apparent in cycle one, which reduces slightly until cycle ten is reached. All following voltage cycles are approximately identical to cycle ten and retain a significant asymmetry demonstrated in the example of cycle 100. As a quantitative measure, we extract the total hysteresis area $A_{H,n}$ normalized to its value at cycle 100 and plot it as a function of the number of cycles for simulations with three different vacancy mobilities (Fig. 7d). All three functions reach $A_{H,n} = 1$ and remain constant after enough cycles. A larger vacancy mobility reduces the number of cycles required to reach the dynamically stable configuration. Consequently, the asymmetry in the I-V characteristics of different Schottky barriers is not a mere volatile feature, as observed in the case of symmetric contacts. Instead, it remains present after a sufficiently large number of cycles reflecting the asymmetric restoring force in equilibrium.

For a more detailed validation, we fitted the model to additional sets of measured I-V curves with various maximum voltages and different extends of asymmetry. The results in Fig. 8 show an overall excellent agreement with the measured data. Only minor deviations are apparent, e.g., in the right hysteresis branch of the asymmetric curves in Fig. 8a and Fig. 8c, with a slightly larger current in the simulation compared to the measurement. All parameters are within the expected range estimated in Section 3 (parameter sets $S_3$-$S_6$, Methods).

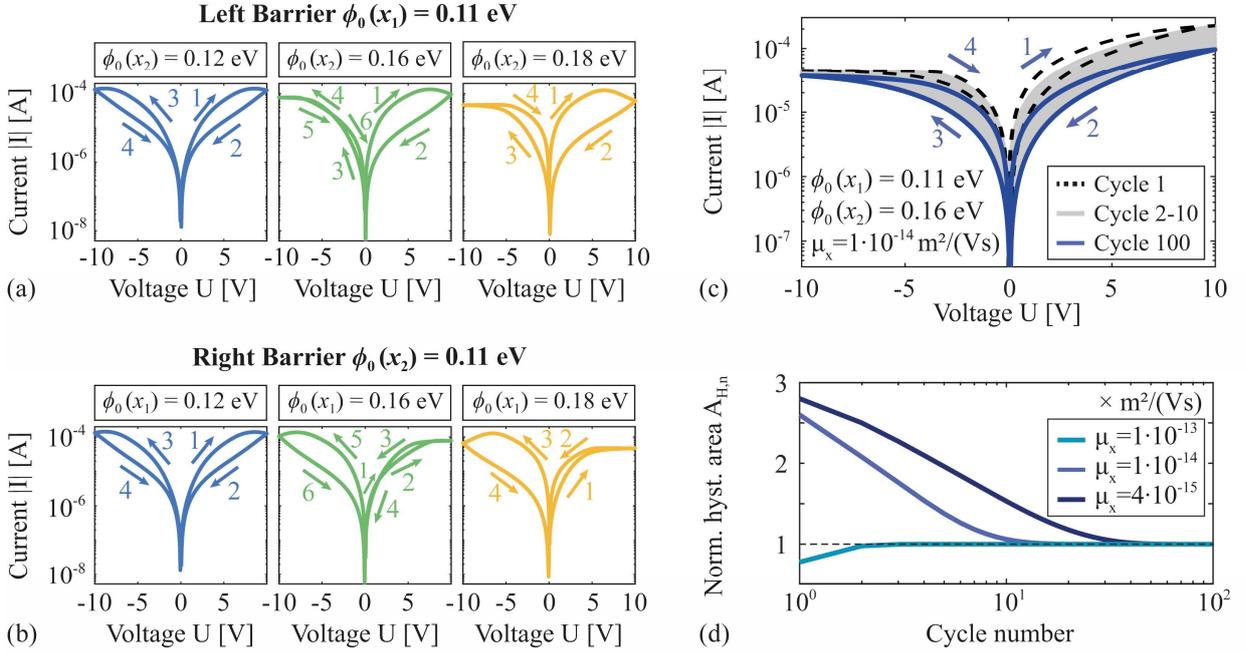

**Fig. 7** Simulations of the second cycle using parameter set $S_2$ (Table 4, Methods) changing only the intrinsic Schottky barriers. **(a)** Current-voltage (I-V) characteristics for three different values of the left barrier $\phi_0(x_1)$ and a fixed value of $\phi_0(x_2) = 0.11$ eV for the right barrier. **(b)** I-V characteristics for three different values of the right barrier $\phi_0(x_2)$ and a fixed value of the left barrier $\phi_0(x_1) = 0.11$ eV. **(c)** I-V characteristics of the first 10 voltage cycles and the $100^{th}$ cycle, with indicated Schottky barriers and vacancy mobility. **(d)** Total hysteresis area as a function of the cycle number normalized to its respective value at cycle 100 and for three different mobilities. The parameter set and the Schottky barriers are identical to those used for the calculations in (c).

## 5 Discussion and Conclusion

We presented a semi-classical charge transport model, which self-consistently couples the drift-diffusion equations for electrons, holes, and charged mobile point defects, to Poisson's equation for the electrostatic potential. We developed and included a formulation for image-charge-induced Schottky barrier lowering (SBL) and solved the system of equations via a finite-volume discretization to analyze the resistive switching mechanism in two-dimensional memristive devices.

The simulations indicate that the hysteresis in all types of I-V curves with and without significant Schottky barriers is caused by the dynamics of mobile charged vacancies, namely the formation and annihilation of a vacancy depletion region, which reduces the conductivity locally and eventually limits the conductivity of the device. The formation and annihilation dynamics of the depletion region are governed by the operating frequency, vacancy mobility, and asymmetry of the two contact potentials. By changing the time and voltage required for the formation of the depletion zone, these parameters crucially determine the hysteresis direction and symmetry of the I-V curve around $U = 0$ V. This switching process explains features of the I-V curve also observed experimentally [34], such as the formation of a maximum at $U < U_{max}$ and the reduction of the maximum current magnitude with increasing mobility. For symmetric contacts, the asymmetry of the I-V curve is volatile and reduces with increasing cycle number before it remains constant.

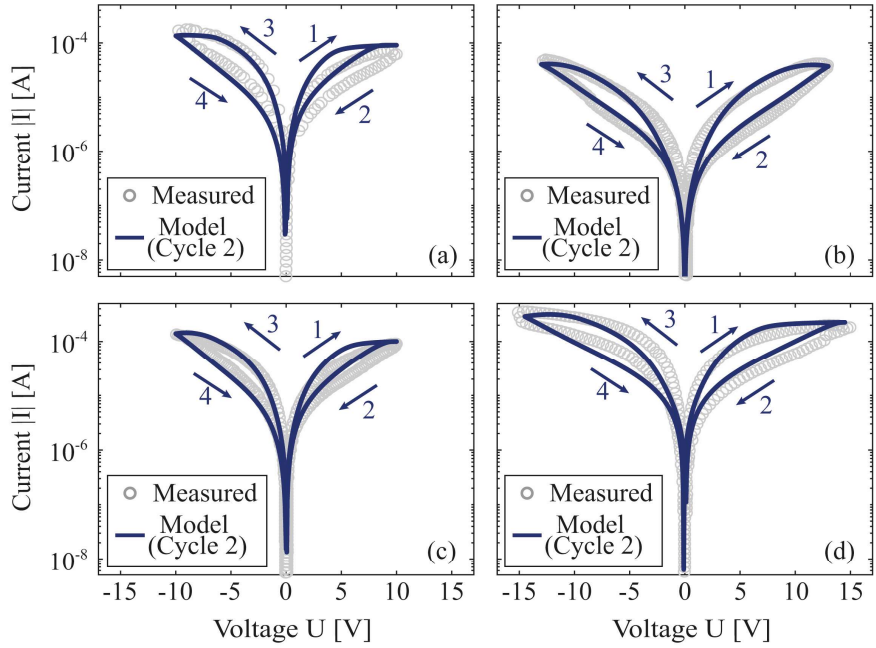

**Fig. 8** Comparison of simulations with measured data of I-V curves of four devices (**a-d**) with various maximum voltages and extends of asymmetry. The measured data are taken from [36], and the simulation parameters are provided in the Methods.

Such an initial cycle-to-cycle instability can be avoided by operating the device in the high-vacancy-mobility range by choosing a corresponding operating frequency (Fig. 5f). Large vacancy mobilities reduce the retention time, which is typically desired for transient memristive devices [87–90] but for most applications of memristive devices as nonvolatile memories, it is undesired [91,92]. Optimizing the cycle-to-cycle stability and retention time implies selecting the narrow range of the sweep rate where the vacancies are sufficiently mobile to occupy their dynamically stable configuration during the first cycle but are still delayed causing a large hysteresis. This is the case at the local maximum in the hysteresis area plot (Fig. 5h), close to the point where the two hysteresis area curves split up.

   A permanent asymmetry in the I-V curve can be obtained by introducing Schottky barriers. We show how minuscule changes in the Schottky barriers (< 0.04 eV) can qualitatively change the I-V characteristic, i.e., reverse the direction and change the hysteresis area and symmetry. Hence, the simulations quantify how deliberate changes can be introduced in the I-V curve for optimizing the device characteristics. Precisely controlling the interface potentials is paramount for fabricating reliable and reproducible devices. Further, we confirmed that SBL represents a crucial contribution to the overall Schottky barrier heights and the slope of the I-V characteristics. SBL reduces the differences between the intrinsic barriers and contributes significantly to their small values in equilibrium. However, while SBL significantly affects the I-V curves, no system configuration was found where SBL determines the hysteresis area. This is because SBL is essentially determined by the projected gradient of the electrostatic potential at the contact interface, and this gradient is by far dominated by the applied voltage and not by the change of the vacancy density at the contacts. While the different hysteresis directions have been previously interpreted as caused by different switching mechanisms [36], our simulations clarify that the exact bulk-limiting mechanism can explain both curves and results in qualitatively different I-V characteristics depending on minor changes in the parameters. While the simulations were performed with $MoS_2$ as a representative example TMDC material, the model is not limited to the drift and diffusion of positively charged sulfur vacancies in $MoS_2$, but it can be applied to other materials and kinds of mobile ionic defects with different charge states.

   The excellent match of the simulations with various experimental I-V curves of lateral $MoS_2$-based devices supports the validity of our results. Also, the model parameters obtained from the fit match very well within the expected range (Section 3). Equilibrium vacancy densities in the simulations are around $n_x \approx 10^{26}$ m$^{-3}$ in good agreement with Auger-electron spectroscopy measurements on the modeled devices [36]. For the electron and hole mobilities, we obtain values of $\mu_{n,p} \approx 1 - 20$ cm$^2$/(Vs), which is well in the reported range of $\mu_{n,p} \approx 0.1 - 200$ cm$^2$/(Vs) [37,69,81–83]. A smaller range of values is obtained for the vacancy mobilities of $\mu_x \approx 1 \cdot 10^{-14} - 1 \cdot 10^{-13}$ m$^2$/(Vs) in the simulations. By using an analytical model [93], these values translate to activation energies of diffusion of $E_a \approx 0.50 - 0.53$ eV (Methods), which matches particularly well with previous estimations for the considered devices of $E_a = 0.6$ eV [36]. *Ab-initio* studies result in larger activation energies of $E_a \approx 0.8 - 2.5$ eV, depending on the microscopic environment, such as the configuration of vacancy clusters and lines [57]. However, the activation energy of diffusion is expected to be further reduced along grain boundaries or interfaces [34]. Hence, our results could support the relevance of such defects for the device performance, consistent with experimental work on MoS$_2$-based lateral devices [34].

The equilibrium Schottky barriers obtained from the simulations vary between approximately 0 eV and 0.17 eV, close to experimental values of $\approx 0.05$ eV [85] for Ti-MoS$_2$ contacts and 0.3 eV from the Schottky-Mott rule [36]. The device-to-device variations of the Schottky barriers, mobilities, and vacancy densities can be caused by microstructure and interface quality variations, reflecting the fabrication tolerances [94,95]. Such variations can also occur in the geometry of the individual devices and explain minor deviations between measurements and simulations. Additionally, the charging and discharging dynamics of trap-states could contribute to the I-V characteristics depending on the defect and interface structure of the individual sample [33,96]. Also, slight deviations from the equilibrium configuration of the system prior to the measurements can introduce an asymmetry in the I-V curves. The overall excellent agreement of the fitted values with experimental data shows that our model can be used beyond a qualitative explanation for the quantitative characterization of the device parameters.

In conclusion, our results strongly corroborate the experimental indications for the relevance of sulfur vacancies in TMDC devices theoretically, and we identified the switching mechanism. Deep insights into the physics of the switching mechanism and the origin of hysteresis are gained as a vital step towards tailoring the device functionality for applications. The results bear general implications for lateral memristive systems and show that mobile ionic defects could dominate their hysteresis. The model and simulation software are made available as powerful tools for the future analysis and design of other semiconductor devices.

# 6  Methods

**Image-charge dielectric constant.** For estimating the image-charge dielectric constant $\varepsilon_i$, the frequency dependency of the in-plane electric permittivity must be known. Measurements on CVD grown monolayers [97] and calculations [98–100] show a strong dispersive behavior at energies of approximately $E > 2$ eV [101] and wavelengths of $\lambda < 900$ nm with real parts of $\text{Re}\{\varepsilon_r^*\} = 8 - 20$ (at $\lambda = 900$ nm) depending on the study [97–100] but generally larger than the corresponding quasi-static values. Further, a*b initio* calculations of the complex permittivity demonstrate that the resonance peaks shift to even higher $E$ and smaller $\lambda$ with increasing number of MoS$_2$ layers until the bulk properties are reached [101]. Therefore, we assume an approximately constant electric permittivity $\varepsilon_r^*$ for energies $E \ll 2$ eV, corresponding to photon frequencies of $f \ll E/h \approx 5 \cdot 10^{14}$ Hz with values possibly slightly larger than the quasi-static value of $\varepsilon_r \approx 10$ for a layer thickness $D = 10 - 15$ nm. Based on the definition of the image-charge dielectric constant $\varepsilon_i \coloneqq \text{Re}\{\varepsilon^*(f_i)\}$ in Section 2.3, we estimate the frequency $f_i = 1/T_i = |\boldsymbol{v}_d|/\Delta x = \mu_n |\boldsymbol{E}|/\Delta x$ ($\boldsymbol{v}_d$: electron drift velocity) by assuming typical values for $\Delta x = 1 - 5$ nm, an electron mobility $\mu_n = 1 - 200$ cm$^2$/Vs and an electric field magnitude $|\boldsymbol{E}| = (0.1 - 1)10^7$ A/m. The resulting frequencies $f_i = 2 \cdot 10^{10} - 2 \cdot 10^{14}$ Hz are at maximum still below the highly dispersive range of $\varepsilon_r^*$, and therefore, we assume $\varepsilon_i \approx \varepsilon_r$.

**Maximum defect density.** We estimate $N_x$ considering hexagonal 2H-MoS$_2$ with an in-plane lattice constant $a = 0.316$ nm and $c = 1.229$ nm [102], which results in a unit cell volume of $V = a^2 c \sin(60°) = 0.1063$ nm$^3 \approx 10^{-28}$ m$^3$. Each unit cell comprises two formula units of MoS$_2$, i.e., two Mo atoms and four S atoms, and is part of two stacked monolayers with a respective distance of $\approx c/2$. With $V$ and the number of S-sites, we estimate a maximum sulfur vacancy density $N_x \approx 4/V \approx 4 \cdot 10^{28}$ m$^{-3}$.

**Activation energy of diffusion and vacancy mobility.** The activation energy of diffusion $E_a$ corresponding to the fitted values of the mobility $\mu_x$ is calculated using the mobility model of Genreith-Schriever [93]. The hopping distance $a_x$ is approximated by the in-plane lattice constant and set to $a_x = 0.316$ nm [102], for the attempt frequency we use a value of $v_0 = 10^{12}$ Hz [103]. We omit the activation entropy of migration $S_a = 0$, and consider the minimum electric field magnitude, i.e., $E_{\min} = 0$, as well as the maximum electric field magnitude of each time step $E_{\max}$ averaged over time. With this range of the electric field, we estimate the error in the activation energy that occurs from assuming a constant vacancy mobility $\mu_x$. The side fraction of vacancies is in good approximation $n_i \coloneqq n_x/N_x \approx 0$. The values for $E_{\max}$ and the resulting $E_a$ are summarized in Table 1 for the two fits from Fig. 4a (data set S$_1$) and Fig. 6a (data set S$_2$).

**Table 1**: Overview of results for the activation energy of diffusion and the parameters extracted from the simulations to calculate the activation energy of diffusion from the vacancy mobilities obtained from the fits in Fig. 4a and Fig. 6a.

| Parameter Name | Expression | Values (S$_1$) | Values (S$_2$) | Unit |
|---|---|---|---|---|
| Mean maximum electric field | $E_{\max}$ | $\approx 1 \cdot 10^{-8}$ | $\approx 3 \cdot 10^{-7}$ | V/m |
| Vacancy mobility | $\mu_x$ | $5 \cdot 10^{-14}$ | $1.15 \cdot 10^{-13}$ | m$^2$/(Vs) |
| Activation energy of diffusion | $E_a$ | $0.53 \pm 0.01$ | $0.503 \pm 0.001$ | eV |

**Summary of model parameters.** In the following, all model parameters and references are summarized. We use a background doping concentration of $C = 10^{21}$ m$^{-3}$ with $z_c = 1$ for all simulations. This concentration value improves numerical stability and is sufficiently small to have no significant influence on the results. Further, a temperature of $T = 300$ K is used. The mobile ionic species x is assumed to behave as n-type dopants [58] with $z_x = +1$ in their ionized state. All material parameters of MoS$_2$ obtained from the literature are summarized in Table 2 together with the respective references. If references are provided for a range of values, the value used from within this range is denoted separately. The geometry parameters are listed in Table 3, and the sample-specific parameters, i.e., the parameter sets S$_1$ and S$_2$, in Table 4.

**Table 2**: Summary of MoS$_2$ material parameters collected from the literature for a layer thickness of $\approx$ 15 nm and comparison with the values used in the simulations. The electron effective mass at rest is denoted by $m_0$, and the effective masses $m_n^*, m_p^*$ are approximated by their bulk values. Values for the fit parameters $\mu_n, \mu_p$ and $\phi_0$ depend on the respective dataset and are provided in Table 4 and for $E_a$ in Table 1.

| Parameter Name | Symbol | Range | Value used | Unit |
|---|---|---|---|---|
| Band gap | $E_g$ | - | 1.3 [66] | eV |
| Electron affinity | $\chi_e$ | 3.7-4.3 [76–80] | 4.0 | eV |
| Relative el. permittivity | $\varepsilon_r$ | - | 10 [74] | 1 |
| Electron/hole effective mass | $m_n^*, m_p^*$ | - | 0.55, 0.71 [62] | $m_0$ |
| Image-charge dielectric constant | $\varepsilon_i$ | $\geq 10$ [97–100] | 10 | 1 |
| Maximum vacancy concentration | $N_x$ | $\leq 4 \cdot 10^{28}$ | $1 \cdot 10^{28}$ | m$^{-3}$ |
| Electron/hole mobility | $\mu_n, \mu_p$ | $0.1 - 200$ [37,69,81–83] | - | cm$^2$/(Vs) |
| Schottky barriers (values for equilibrium) | $\phi(x)$ | 0.05 [85], 0.3 [36] | - | eV |
| Activation energy of diffusion | $E_a$ | 0.6 [36], $0.8 - 2.5$ [57] | - | eV |

**Table 3**: Summary of geometry parameters used for the simulations of the device in [36]. The two values provided for the channel length $L$ are used together with the parameter sets S$_1$-S$_6$ (Table 4, Table 5).

| Parameter Name | Expression | Values used | Unit |
|---|---|---|---|
| Channel length | $L$ | 1 (S$_1$, S$_3$-S$_5$), 2 (S$_2$, S$_6$) | |
| Channel width | $W$ | 10 | μm |
| Channel thickness | $D$ | 0.015 | |

**Table 4:** Sample-specific parameter sets S$_1$ and S$_2$ obtained from the fit of the simulation to the experimental data in Fig. 4a (S$_1$) and Fig. 6a (S$_2$). For all fits a sweep rate of $f = 5$ V/s was used.

| Parameter Name | Symbol | S$_1$ | S$_2$ | Units |
|---|---|---|---|---|
| Left Schottky barrier | $\phi_0(x_1)$ | $1 \cdot 10^{-3}$ | 0.144 | eV |
| Right Schottky barrier | $\phi_0(x_2)$ | | 0.110 | eV |
| Electron/hole mobility | $\mu_n, \mu_p$ | $2.5 \cdot 10^{-4}$ | $2.15 \cdot 10^{-3}$ | m$^2$/(Vs) |
| Vacancy mobility | $\mu_x$ | $5 \cdot 10^{-14}$ | $1.15 \cdot 10^{-13}$ | m$^2$/(Vs) |
| Intrinsic defect energy | $E_{x,0}$ | -4.32 | -4.33 | eV |
| Voltage amplitude | $U_{max}$ | 13 | 10 | V |

**Table 5:** Parameters obtained from the fit of the model to the experimental data in Fig. 8a-d. All fits were performed with a sweep rate of $f = 5$ V/s.

| Parameter Name | Symbol | S$_3$ (Fig. 8a) | S$_4$ (Fig. 8b) | S$_5$ (Fig. 8c) | S$_6$ (Fig. 8d) | Units |
|---|---|---|---|---|---|---|
| Left Schottky barrier | $\phi_0(x_1)$ | 0.167 | $1 \cdot 10^{-3}$ | 0.152 | 0.155 | eV |
| Right Schottky barrier | $\phi_0(x_2)$ | 0.148 | | 0.001 | 0.115 | eV |
| Electron/hole mobility | $\mu_n, \mu_p$ | $1.3 \cdot 10^{-3}$ | $3.6 \cdot 10^{-4}$ | $1.5 \cdot 10^{-3}$ | $6 \cdot 10^{-4}$ | m$^2$/(Vs) |
| Vacancy mobility | $\mu_x$ | $8.5 \cdot 10^{-14}$ | $6 \cdot 10^{-14}$ | $7 \cdot 10^{-14}$ | $3.4 \cdot 10^{-14}$ | m$^2$/(Vs) |
| Intrinsic defect energy | $E_{x,0}$ | -4.3 | -4.32 | -4.32 | -4.28 | eV |
| Voltage amplitude | $U_{max}$ | 10 | 13 | 10 | 14.5 | V |

**Funding**
The research is funded by the Carl-Zeiss Foundation via the Project Memwerk, the Deutsche Forschungsgemeinschaft (DFG, German Research Foundation)– Project-ID 434434223 – SFB 1461, and the Leibniz Competition 2020.

**Competing interests**
The authors declare that they have no competing interests.

**Availability of data and materials**
The datasets used and analyzed during the current study are available from the corresponding author upon reasonable request. The simulated data sets can be generated with the open-source software tool ChargeTransport.jl (https://github.com/PatricioFarrell/ChargeTransport.jl).